\documentclass{article}
\pdfoutput=1

\usepackage[english]{babel}
\usepackage[utf8]{inputenc}
\usepackage[T1]{fontenc}

\usepackage[a4paper,top=2.5cm,bottom=3cm,left=2.5cm,right=2.5cm,marginparwidth=1.75cm]{geometry}

\usepackage{amsmath}
\usepackage[round]{natbib}
\usepackage{graphicx}
\usepackage[colorlinks=true, allcolors=blue]{hyperref}
\usepackage[detect-weight,retain-explicit-plus]{siunitx} %
\usepackage{makecell}

\usepackage[yyyymmdd]{datetime}
\usepackage{rotating}
\usepackage{geometry}
\usepackage{pdflscape}
\usepackage{tablefootnote}

\title{Binaural Audio Rendering in the Spherical Harmonic Domain: \linebreak
A Summary of the Mathematics and Its Pitfalls}
\author{Jens Ahrens}
\date{Technical note\footnote{Find more technical notes at \url{http://www.ta.chalmers.se/education/texts-on-acoustics/}}~~v.~2, Feb.~2022\\[1ex] 
Chalmers University of Technology\\[1ex] 
\texttt{jens.ahrens@chalmers.se}}

\newcommand{\e}{\mathrm{e}}
\renewcommand{\d}{\mathrm{d}}
\renewcommand{\i}{\mathrm{i}}

\begin{document}
\maketitle

\begin{abstract}
The present document reviews the mathematics behind binaural rendering of sound fields that are available as spherical harmonic expansion coefficients. This process is also known as binaural ambisonic decoding. We highlight that the details entail some amount peculiarity so that one has to be well aware of the precise definitions that are chosen for some of the involved quantities to obtain a consistent formulation. We also discuss what sets of definitions produce ambisonic signals that are compatible with the most common software tools that are available. 
\end{abstract}

\section{Introduction}

Any sound presentation method that uses loudspeakers can also be used for binaural rendering simply by using head-related transfer functions (HRTFs) as virtual loudspeakers. This approach has also been used in the context of ambisonics in order to achieve what is often referred to as \emph{binaural decoding} in~\citep{jot1998approaches,Wiggins:2001,Noisternig:2003} and in a more elaborate way in~\citep{Duraiswami:AES2005}. The ambisonic signals that are being decoded originate usually from a microphone array. Very typical are spherical microphone arrays with a rigid baffle~\citep{Zotter:book2019}.

More recent approaches to binaural rendering of such ambisonic sound scenes -- or spherical harmonic (SH) representations of sound scenes -- use a different formulation that does not employ discrete virtual loudspeakers but a continuous distribution thereof~\citep{Menzies:JASA2007,Rafaely:JASA2010,Bernschutz:PhD2016}. One also speaks of binaural decoding or binaural rendering \emph{in the SH domain}. Conceptually, these methods may be summarized as follows: The SH representation of the captured sound field provides information on what the sound field is; the HRTFs represent how the human physiology influences a sound field on its way to the ear canal entrances. The methods then establish a mathematical formulation that 'applies' the HRTFs on the captured sound field. The result is the signals that would arise at the ears of a listener when he/she were exposed to the sound field that is represented by the SH signals. The listener is virtually placed into the sound field.

We review this more recent rendering method in this document in the variant proposed in~\citep{Rafaely:JASA2010}. The purpose of this document is summarizing the mathematical formulation in one single place with consistent nomenclature and definitions of the building blocks. We highlight what alternative definitions of some of the building blocks are commonly used, and how the presented formulation needs to be adapted if those are being employed. Mastering this will allow for arbitrarily combining the work from different authors while maintaining a consistent formulation.

\section{Definition of the Fourier Transform Over Time}\label{sec:def_fourier}

In the following, we assume that the forward and inverse Fourier transforms over time are defined as
\begin{equation}\label{eq:cft}
     S(\omega)  = \int\displaylimits_{-\infty}^{\infty} s(t)  \ \e^{-\i \omega t} \mathrm{d}t 
\end{equation}
and
\begin{equation}\label{eq:cift}
     s(t) = \frac{1}{2\pi} \int\displaylimits_{-\infty}^{\infty} S(\omega)  \ \e^{\i \omega t} \mathrm{d}\omega \, ,
\end{equation}
where $s(t)$ is the time-domain representation of our signal under consideration and $S(\omega)$ its frequency-domain representation. Note that the algebraic sign of the exponent of the complex exponential is the aspect of importance. Above definitions are compatible with the definition of the discrete Fourier transform in MATLAB\footnote{Note that~\citep[Eq.~(1.6)]{Williams:book1999}, for example, uses a different sign convention.}.\\

The above definitions of the forward and inverse Fourier transform over time determine that the spherical Hankel function of second kind $h_n^{(2)}(\cdot)$ denotes outward propagating waves. Refer to~\citep{Ahrens:sign_conventions} for an in-depth discussion of the matter of sign conventions with waves.

\section{Spherical Harmonic Decomposition by Means of Spherical Microphone Arrays}\label{sec:sh_decomposition}

Any square-integrable function $S(\beta, \alpha)$ that is defined on the surface of a sphere as a function of the spatial coordinates colatitude\footnote{The colatitude is also referred to as \emph{polar angle} or \emph{zenith angle}. Note that it is not elevation!} $\beta$ and azimuth $\alpha$ can be expressed as a series of surface spherical harmonics $Y_{n,m}(\beta, \alpha)$ as~\mbox{\citep{Gumerov:book2005}}
\begin{equation}\label{eq:sh_decomp_basic}
    S(\beta, \alpha) = 
    \sum_{n=0}^\infty \sum_{m=-n}^n\! \mathring{S}_{n,m} \,Y_{n,m}(\beta, \alpha) \ .
\end{equation}
$\mathring{S}_{n,m}$ are termed the SH~coefficients of $S(\beta, \alpha)$. 

We assume that $Y_{n,m}(\beta, \alpha)$ is defined as~\citep[Eq.~(2.1.59)]{Gumerov:book2005}
\begin{equation}\label{eq:sh_definition}
    Y_{n,m}(\beta, \alpha)= (-1)^m\, \sqrt{\frac{2n{+}1}{4\pi} \frac{(n{-}|m|)!}{(n{+}|m|)!}}\,P_n^{|{m}|}(\cos\beta)\,\,\e^{\i m \alpha}  ,
\end{equation}
in which $P_n^{|{m}|}(\cdot)$ are the associated Legendre functions\footnote{Associated Legendre functions are used with different normalizations. We assume that $P_n^{m}(\mu)$ is defined via the following Rodriguez formula~\citep[Eq.~(2.1.20)-(2.1.21)]{Gumerov:book2005}:
\begin{equation}
    P_n^{m}(\mu) = (-1)^m(1-\mu^2)^{m/2} \frac{\d^m}{\d \mu^m} P_n(\mu),  \ \ \forall n\geq 0, \ m\geq m \ ,
\end{equation}
with
\begin{equation}
    P_n(\mu) = \frac{1}{2^n n!} \frac{\d^n}{\d \mu^n} (\mu^2 - 1),  \ \ \forall n\geq 0 \ .
\end{equation}
This also the definition that MATLAB and SciPy use.
}, $\i$ is the imaginary unit\@. Note that there are several alternative viable definitions of $Y_{n,m}(\cdot)$ including purely real ones. We highlight at this point that the formulation of some of the results derived below depends to some extent on the definition of $Y_{n,m}(\cdot)$. We refer the reader to Sec.~\ref{sec:definition_of_Y_nm} for a discussion of this.\hfill \break

The coefficients $\mathring{S}_{n,m}$ are obtained from
\begin{equation}\label{eq:direct_comp_sh_coeff_0}
     \mathring{S}_{n,m} 
    = \oint\displaylimits_{\Omega \in S^2}\! S(\beta, \alpha)\,\,Y_{n,m}(\beta, \alpha)^*\,\d\Omega \ .
\end{equation}
The asterisk~$^*$ denotes complex conjugation. $\Omega$ is an infinitesimal surface element of the unit sphere $S^2$ (Sorry for using $S$ here again. It's the standard symbol for the unit sphere.). The integration over the surface of the unit sphere in~\eqref{eq:direct_comp_sh_coeff_0} is explicitly performed as
\begin{equation}\label{eq:direct_comp_sh_coeff_1}
     \mathring{S}_{n,m} 
    = \int\displaylimits_0^{2\pi} \! \int\displaylimits_0^{\pi} S(\beta, \alpha)\,\,Y_{n,m}(\beta, \alpha)^*\cos \beta \, \d \beta \d \alpha \ .  
\end{equation}
In the remainder, we assume that the quantity $S(\cdot)$ is frequency dependent, i.e., $S(\beta, \alpha) = S(\beta, \alpha, \omega)$ and therefore $\mathring{S}_{n,m} = \mathring{S}_{n,m}(\omega)$ where $\omega{\,=\,}2\pi f$~is the radian frequency in~\si{rad/s}\@, $f$~is the frequency in~\si{Hz}. If $S(\cdot)$ is an interior sound pressure field $S(\beta, \alpha, r, \omega)$, whereby $r$ is the radial distance from the coordinate origin, it can be represented as~\mbox{\citep{Gumerov:book2005}}
\begin{equation}\label{eq:sh_decomp_free-field}
    S(\vec{x}, \omega) = S(\beta, \alpha, r, \omega) = 
    \sum_{n=0}^\infty \sum_{m=-n}^n\! \mathring{S}_{n,m}(r, \omega)  \,\,Y_{n,m}(\beta, \alpha) \ ,
\end{equation}
with
\begin{equation}\label{eq:ring_vs_breve}
    \mathring{S}_{n,m}(r, \omega) = \breve{S}_{n,m}(\omega)\,\,j_n(r, \omega) \ .
\end{equation}
$j_n(\cdot)$ are the spherical Bessel functions. 

The goal of the following endeavor is measuring the coefficients $\breve{S}_{n,m}(\omega)$ of a given sound field. Pressure microphones that are distributed over the surface of a rigid spherical scattering object were shown to be a suitable apparatus for this~\citep{Meyer:ICASSP2002,Abhayapala:ICASSP2002,Rafaely:TASPL2005}. In the following, we assume that $\e^{-\i \vec{\mathbf{k}}_\text{pw}^T \vec{\mathbf{x}}}$ represents a plane wave that propagates in the direction into which $\vec{\mathbf{k}}_\text{pw}$ points so that spherical Bessel functions of second kind $h_n^{(2)}\!\left(\cdot \right)$ denote outward propagating waves (cf.~also Sec.~\ref{sec:sign_conventions}).

A sound pressure field $S^\text{\;\!surf}(\beta, \alpha, R, \omega)$ on the surface of a rigid spherical scatterer of radius~$R$ that is centered at the coordinate origin is given by~\mbox{\citep[Eq.~(3.1.1)]{Gumerov:book2005}}
\begin{equation}\label{eq:sh_decomp_sphere}
    S^\text{\;\!surf}(\beta, \alpha, R, \omega) = 
    \sum_{n=0}^\infty \sum_{m=-n}^n\! \mathring{S}_{n,m}^\text{\;\!surf}(R, \omega)  \,\,Y_{n,m}(\beta, \alpha) \ ,
\end{equation}
with
\begin{equation}\label{eq:mathring_vs_breve}
    \mathring{S}_{n,m}^\text{\;\!surf}(R, \omega) = \breve{S}_{n,m}(\omega)\,\,b_n(R, \omega) \ .
\end{equation}
$b_n(\cdot)$ is given by~\citep[Eq.~(4.2.10) and (4.2.13)]{Gumerov:book2005}
\begin{eqnarray}\label{eq:radial_filter_3d}
    b_n\!\left( R, \omega\right) & = &
    j_n\!\left( \omega \frac{R}{c} \right) - \frac{j_n^{\;\!\prime}\!\left( \omega \frac{R}{c} \right)}{h_n^{\prime\;\!(2)}\!\left( \omega \frac{R}{c} \right)}\,h_n^{(2)}\!\!\left( \omega \frac{R}{c} \right)\\ \label{eq:radial_filter_3d_prime}
    & = & -\frac{\i}{\left(\omega\frac{R}{c}\right)^2} \,\frac{1}{h_n^{\prime\;\!(2)}\!\!\left( \omega \frac{R}{c} \right)} \ .
\end{eqnarray}
$c$~is the speed of sound in~\si{m/s}, and $h_n^{\prime\;\!(2)}\!(\cdot)$~denotes the derivative of the $n$th~order spherical Hankel function of second kind with respect to the argument, $j_n^{\;\!\prime}\!\left( \cdot \right)$ the derivative of the $n$th~order spherical Bessel function with respect to the argument.

$\mathring{S}_{n,m}^\text{\;\!surf}(R, \omega)$ are the SH~coefficients of the sound pressure on the surface of the spherical scatterer with radius~$R$\@. $\breve{S}_{n,m}(\omega)$ are the SH~coefficients --~and thereby a complete representation~-- of the incident sound field with the effect of the scatterer removed. We are not aware that there is an established terminology that allows for differentiating between the SH coefficients $\mathring{S}_{n,m}(\cdot)$ that we mark with a circle and the SH coefficients $\breve{S}_{n,m}(\cdot)$ that we mark with a breve. We have to use the mathematical notation for this.

Let us distribute pressure microphones on the surface of the spherical scatterer and thereby establish a spherical microphone array (SMA). The computation of~$\breve{S}_{n,m}(\omega)$ from the microphone signals $S^\text{\;\!surf}(\beta, \alpha, R, \omega)$ can be performed via~\citep{Rafaely:TASPL2005}
\begin{equation}\label{eq:direct_comp_sh_coeff}
    \mathring{S}_{n,m}^\text{\;\!surf}(R, \omega) 
    = \oint\displaylimits_{\Omega \in S^2}\! S^\text{\;\!surf}(\beta, \alpha, R, \omega)\,\,Y_{n,m}(\beta, \alpha)^*\,\d\Omega 
\end{equation}
and
\begin{equation}\label{eq:breve_from_ring}
    \breve{S}_{n,m}(\omega) = \mathring{S}_{n,m}^\text{\;\!surf}(R, \omega)\,\, b_n^{-1}(R, \omega) \ ,
\end{equation}
or equivalently,
\begin{equation}\label{eq:s_breve_from_s}
    \breve{S}_{n,m}(\omega)
    = b_n^{-1}(R, \omega)\, \oint\displaylimits_{\Omega \in S^2}\! S^\text{\;\!surf}(\beta, \alpha, R, \omega) \,\,Y_{n,m}(\beta, \alpha)^*\,\d\Omega \ .
\end{equation}
$b_n^{-1}(R, \omega)$ is termed \emph{radial~filters} in the SMA literature. These filters exhibit impractically high gains at low frequencies at high orders (because $b_n(R, \omega)$ tends to~0 there) so that they require regularization. The effect of this is well documented in the SMA literature~\citep{Moreau:AES2006,Bernschutz:PhD2016}.


In practical implementations, the integrals in~\eqref{eq:direct_comp_sh_coeff} and~\eqref{eq:s_breve_from_s} are discretized because a continuous layer of pressure microphones cannot be implemented but a finite set of discrete microphones has to be used. This bounds the maximum order~$n$ that can be extracted to~$n\leq N$ because the orthogonality of $Y_{n,m}$ is maintained only then\@. One speaks of an $N$th~order decomposition.

\section{Binaural Rendering}\label{sec:binaural_rendering}

This section summarizes how a sound field whose SH coefficients $\breve{S}_{n,m}(\omega)$ are known can be auralized -- i.e.~rendered/decoded -- binaurally. We define an HRTF $H(\phi, \theta, \omega)$ as the ear response to a plane wave that carries a time-domain impulse and that propagates into the direction $(\phi, \theta)$, whereby $\phi$ is the colatitude and $\theta$ is the azimuth in a spherical coordinate system\footnote{HRTFs do indeed exhibit some amount of distance dependence. But this is only relevant at distances of, say, \SI{1}{m} or shorter (cf.,~for example,~\citep{Wierstorf:AES2011}). HRTFs that where measured at farther distances can be interpreted as plane-wave HRTFs.}. 

Any interior sound field $S(\vec{x}, \omega)$ can be represented by its \emph{Herglotz kernel} or \emph{signature function} $\bar{S}(\phi, \theta, \omega)$ as~\citep[Sec.~1.2.1.7, Eq.~(72)]{kress:2001}, \citep[Eq.~(7.1.27)]{Gumerov:book2005} 
\begin{equation}\label{eq:signature_function}
    S(\vec{x}, \omega) = \oint\displaylimits_{\Omega \in S^2} \bar{S}(\phi, \theta, \omega) \, \mathrm{e}^{- \mathrm{i} \vec{k}(\phi, \theta, \omega)^T \vec{x}} \ \mathrm{d} \Omega \, ,
\end{equation}
whereby $\Omega = \Omega(\phi, \theta)$ is an infinitesimal surface element of the unit sphere $S^2$, $\vec{x} = [x, y, z]^T$ is a location in a Cartesian coordinate system, and $\vec{k}(\phi, \theta, \omega) = \frac{\omega}{c} [\cos \theta \sin \phi, \sin \theta \sin \phi, \cos \phi]^T$ is the wave vector. In other words, $S(\vec{x}, \omega)$ is represented by a continuum of propagating plane waves, and $\bar{S}(\phi, \theta, \omega)$ are the coefficients of the plane waves\footnote{Despite using propagating plane waves as basis functions, the representation~\eqref{eq:signature_function} is a general representation that applies to \emph{any} interior sound field~\citep[Sec.~1.2.1.7]{kress:2001}.}.

It is presented in~\citep[Eq.~(7.1.26)-(7.1.28)]{Gumerov:book2005} (and more explicitly in~\citep[App.~E.4.2]{Ahrens:2012}) that $\bar{S}(\phi, \theta, \omega)$ is related to the SH coefficients $\breve{S}_{n,m}(\omega)$ of $S(\vec{x}, \omega)$ by 
\begin{equation}\label{eq:S_bar_to_breve_S}
    \bar{S}(\phi, \theta, \omega) = \sum_{n=0}^\infty \sum_{m = -n}^n \frac{1}{4\pi\mathrm{i}^{-n}} \, \breve{S}_{n,m}(\omega) Y_{n,m} (\phi, \theta) \ .
\end{equation}

In order to compute the signal $B(\omega)$ that arises at the ears of a listener due to exposure of the listener to $S(\vec{x}, \omega)$, we can employ the representation~\eqref{eq:signature_function}. We recall that an HRTF represents the ear response to a plane wave. Instead of integrating over plane waves like in~\eqref{eq:signature_function}, we integrate over the HRTFs $H(\phi, \theta, \omega)$ and weight each HRTF with the strength $\bar{S}(\phi, \theta, \omega)$ of the corresponding plane wave component as~\citep{Menzies:JASA2007}
\begin{eqnarray} \nonumber
    B(\omega) &=& \oint\displaylimits_{\Omega \in S^2} \bar{S}(\phi, \theta, \omega) \, H(\phi, \theta, \omega) \ \mathrm{d} \Omega \\ \nonumber
              &=& \oint\displaylimits_{\Omega \in S^2} \sum_n \sum_m \frac{1}{4\pi\mathrm{i}^{-n}}  \breve{S}_{n,m}(\omega) Y_{n,m} (\phi, \theta) \times \, \sum_{n'} \sum_{m'} \mathring{H}_{n',m'}(\omega) Y_{n',m'} (\phi, \theta) \ \mathrm{d} \Omega \\ \label{eq:binaural_rendering_ansatz}
              &=&  \sum_n \sum_m \frac{1}{4\pi\mathrm{i}^{-n}}  \breve{S}_{n,m}(\omega)  \sum_{n'} \sum_{m'} \mathring{H}_{n',m'}(\omega) \times \!\! \oint\displaylimits_{\Omega \in S^2} \! Y_{n,m} (\phi, \theta)\ Y_{n',m'} (\phi, \theta) \ \mathrm{d} \Omega  \ .
\end{eqnarray}
The basis functions $Y_{n,m} (\phi, \theta)$ are orthonormal, which means that 
\begin{equation}\label{eq:orthonormality}
     \oint\displaylimits_{\Omega \in S^2} Y_{n,m} (\phi, \theta)^* \, Y_{n',m'} (\phi, \theta)  \ \mathrm{d} \Omega = \delta_{n,n'} \delta_{m,m'} \, ,
\end{equation}
whereby $\delta$ denotes the Kornecker delta. Noting that $Y_{n,-m} (\cdot) = Y_{n,m} (\cdot)^*$ for the present definition of $Y_{n,m} (\cdot)$ (Eq.~\eqref{eq:sh_definition}), we reformulate~\eqref{eq:orthonormality} as
\begin{equation}\label{eq:orthonormality2}
     \oint\displaylimits_{\Omega \in S^2} Y_{n,-m} (\phi, \theta) \, Y_{n',m'} (\phi, \theta)  \ \mathrm{d} \Omega = \delta_{n,n'} \delta_{m,m'} \, .
\end{equation}

In order to apply~\eqref{eq:orthonormality2} in~\eqref{eq:binaural_rendering_ansatz}, we replace $m$ in~\eqref{eq:binaural_rendering_ansatz} with $-m$ without loss of generality. This leads finally to
\begin{equation}\label{eq:binaural_rendering_final}
    B(\omega) = \sum_{n=0}^\infty \sum_{m = -n}^n
    \frac{1}{4\pi\mathrm{i}^{-n}}  \breve{S}_{n,-m}(\omega) \ \mathring{H}_{n,m}(\omega)  \, .
\end{equation}
Eq.~\eqref{eq:binaural_rendering_final} may be termed the \emph{binaural rendering equation}. A variant of it was initially published in~\citep{Rafaely:JASA2010}. It describes how the left and right ear signals $B^\text{L,R}(\omega)$ are obtained from the SH coefficients $\breve{S}_{n,-m}(\omega)$ of the sound field and the SH coefficients $\mathring{H}_{n,m}^\text{L,R}(\omega)$ of the left-ear and right-ear HRTFs. Recall~\eqref{eq:sh_decomp_basic} and \eqref{eq:ring_vs_breve}-\eqref{eq:mathring_vs_breve} to appreciate the difference between the coefficients marked with a breve ($\breve{X}_{n,m}$) and those marked with a ring ($\mathring{X}_{n,m}$).

\section{Peculiarities}\label{sec:peculiarities}

\subsection{Sign Conventions}\label{sec:sign_conventions}

As indicated in Sec.~\ref{sec:def_fourier}, the choice of the signs in the exponents of the complex exponential in the time and the spatial Fourier transforms determine what the propagation direction of a wave is that is represented by a given mathematical expression. The chosen sign convention does indeed affect the mathematical expressions that are used in the present document. We refer the reader to~\citep{Ahrens:sign_conventions} for more information. 

\subsection{Definition of the HRTFs}\label{sec:definition_of_hrtfs}

HRTFs are usually defined as the ear response to sound impinging from a given source position or direction. This is also how the data are stored. Yet, as per the definitions above, $\mathring{H}_{n,m}(\omega)$ are the SH coefficients under the interpretation that an HRTF is the ear response to a plane wave that propagates into a given direction $(\phi, \theta)$. Otherwise, Eq.~\eqref{eq:binaural_rendering_ansatz} would not be applicable.

This means that we have to convert the coordinates of the sound incidence directions that are usually part of the HRTF data into propagation directions before being able to use~\eqref{eq:direct_comp_sh_coeff_0} for expanding $H(\phi_\textnormal{prop.}, \theta_\textnormal{prop.}, \omega)$ into $\mathring{H}_{n,m}(\omega)$. This is conversion is straightforward:
\begin{eqnarray} \nonumber
    \phi_\textnormal{prop.} &=& \pi - \phi_\textnormal{incid.}\\
    \theta_\textnormal{prop.} &=& \pi + \theta_\textnormal{incid.} \ .
\end{eqnarray}
Recall that $\phi$ denotes colatitude, and $\theta$ denotes azimuth.\\

Let us quickly look at this matter from a mathematical point of view. Replacing $\alpha$ with $\pi+\alpha$ and $\beta$ with $\pi - \beta$ in the SH transformation in~\eqref{eq:direct_comp_sh_coeff} means for the basis function that
\begin{equation}\label{eq:inc_vs_prop}
    Y_{n,m}(\pi - \beta, \pi + \alpha)^* = (-1)^n (-1)^m (-1)^{-m} Y_{n,m}(\beta, \alpha)^*
\end{equation}
We exploited~\cite[Eq.~(2.1.46)]{Gumerov:book2005} in~\eqref{eq:inc_vs_prop}. The terms $(-1)^m$ and $(-1)^{-m}$ cancel out. This means that instead of obtaining the coefficients $\mathring{H}_{n,m}(\omega)$, we obtain $(-1)^n \mathring{H}_{n,m}(\omega)$. This in turn means that we can compensate for such a confusion of the incidence and propagation directions by applying a factor $(-1)^n$ to the ambisonic signals. As a side note, applying this to~\eqref{eq:binaural_rendering_final} leads to 
\begin{equation}\label{eq:binaural_rendering_final_2}
    B'(\omega) = \sum_{n=0}^\infty \sum_{m = -n}^n
    \frac{1}{4\pi\mathrm{i}^{n}}  \breve{S}_{n,-m}(\omega) \ \mathring{H}_{n,m}(\omega)  \, .
\end{equation}
Replacing $\i^{-n}$ with $\i^n$ (and vice versa) therefore also compensates for this confusion.

\subsection{Definition of the Spherical Harmonics \texorpdfstring{$Y_{n,m} (\phi, \theta)$}{Y}}
\label{sec:definition_of_Y_nm}

Sec.~\ref{sec:sh_decomposition} stays entirely viable with any conceivable definition of $Y_{n,m} (\phi, \theta)$. Using a real valued definition simply makes the complex conjugation that is performed in some equations redundant.\hfill \break

Let us reconsider Sec.~\ref{sec:binaural_rendering} on the example of the definition used in~\citep[Eq.~(6.20)]{Williams:book1999}, which reads
\begin{equation}\label{eq:sh_definition_2}
    Y_{n,m}(\beta, \alpha) = \sqrt{\frac{2n{+}1}{4\pi} \frac{(n{-}m)!}{(n{+}m)!}}\,P_n^{{m}}(\cos\beta)\,\,\e^{\i m \alpha}  \, .
\end{equation}
Compare~\eqref{eq:sh_definition_2} of~\eqref{eq:sh_definition} and appreciate the differences. Here's the crux: While the previous definition~\eqref{eq:sh_definition} leads to $Y_{n,-m} (\cdot) = Y_{n,m} (\cdot)^*$, the present definition~\eqref{eq:sh_definition_2} leads to~\citep[Eq.~(6.44)]{Williams:book1999}
\begin{equation}
    Y_{n,-m} (\cdot) = (-1)^m \, Y_{n,m} (\cdot)^*  \ . 
\end{equation}
As a consequence, we need to add a factor of $(-1)^m$ to~\eqref{eq:orthonormality2} and \eqref{eq:binaural_rendering_final} if~\eqref{eq:sh_definition_2} is employed~\citep{Andersson:MSc2017}.

Missing out on this factor of $(-1)^m$ does not make the binaural signals unusable. It rather rotates the scene by \SI{180}{\degree} along the azimuth, because $(-1)^m = \e^{\i m \pi}$, which one can detect easily and correct for accordingly.\hfill \break

\noindent When using a real definition of $Y_{n,m}(\beta, \alpha)$\footnote{Such as~\cite[Sec.~18.4]{Whittaker:book1963}
\begin{equation}\label{eq:sh_definition_real}
    Y_{n,m}(\beta, \alpha) = (-1)^m\, \sqrt{\frac{2n{+}1}{4\pi} \frac{(n{-}|m|)!}{(n{+}|m|)!}}\,P_n^{|{m}|}(\cos\beta)
\begin{cases}
 \sqrt{2} \sin{|m| \alpha}, \ \forall m < 0\\
 \ \ \ \  \ \ \ 1 \ \ \ \ \ \ \, , \ \forall m = 0\\
 \sqrt{2} \cos{|m| \alpha}, \ \forall m > 0
\end{cases} \ ,
\end{equation}
which corresponds to the definition that is used in the N3D ambisonic format~\citep{Nachbar:AmbiX2011}. It is also identical to the definition that is used in Politis's \texttt{getSH.m} MATLAB function when called with the argument \texttt{'real'}~\citep{Politis:SH_transform}.}, the orthonormality relation~\eqref{eq:orthonormality} reads 
\begin{equation}\label{eq:orthonormality3}
     \oint\displaylimits_{\Omega \in S^2} Y_{n,m} (\phi, \theta) \, Y_{n',m'} (\phi, \theta)  \ \mathrm{d} \Omega = \delta_{n,n'} \delta_{m,m'} \, .
\end{equation}
Exploiting this in~\eqref{eq:binaural_rendering_ansatz} turns the rendering equation~\eqref{eq:binaural_rendering_final} into
\begin{equation}\label{eq:binaural_rendering_final_real}
    B(\omega) = \sum_{n=0}^\infty \sum_{m = -n}^n
    \frac{1}{4\pi\mathrm{i}^{-n}}  \breve{S}_{n,m}(\omega) \ \mathring{H}_{n,m}(\omega)  \, .
\end{equation}
%

\subsection{Propagation Direction of a Plane Wave Vs.~Incidence Direction}\label{sec:propagation_direction}

Some authors prefer to define plane waves not by their propagation direction but by their direction of incidence (see, for example,~\citep{Rafaely:JASA2010}). This requires a lot of modifications of the present formulation as, for example,~\eqref{eq:signature_function} changes and also the whole spiel around the HRTFs that is discussed in Sec.~\ref{sec:definition_of_hrtfs} is exactly the other way round. Not least does this relate to the general topic of sign conventions (of the exponent of the complex exponential that described the spatial part of a sound field), which is in itself rather convoluted (we refer the reader to~\citep{Ahrens:sign_conventions} for more information on that.). The matter of propagation direction vs.~incidence direction is discussed further in Sec.~\ref{sec:ambisonics}.


\subsection{Ambisonics}\label{sec:ambisonics}

The formulation presented in Sec.~\ref{sec:sh_decomposition} is heavily inspired by~\citep{Gumerov:book2005}. It can be considered more of a mathematical point of view. When using the solutions in practice, it is worth adopting also the physical perspective. One way doing this is defining the coefficients $\breve{S}_{n,m}(\omega)$ via
\begin{equation}\label{eq:sh_basic_ambi}
    S(\vec{x}, \omega) = S(\beta, \alpha, r, \omega) = 
    \sum_{n=0}^\infty \sum_{m=-n}^n\! 4\pi\mathrm{i}^{-n} \breve{S}_{n,m}(\omega)\,\,j_n(r, \omega) \,\,Y_{n,m}(\beta, \alpha)
\end{equation}
rather than via~\eqref{eq:sh_decomp_free-field} and~\eqref{eq:ring_vs_breve}. The main motivation for the definition~\eqref{eq:sh_basic_ambi} is that it yields time-domain signals $\breve{s}_{n,m}(t)$ that are real (for real sound pressures $s(\vec{x}, t)$) when using a real-valued definition of $Y_{n,m}(\cdot)$ so that they can be stored as multichannel audio signals\footnote{A secondary aspect is that this formulation allows for interpreting $\breve{S}_{n,m}(\omega)$ as a plane wave representation~\citep[Ch.~2.4 and Eq.~(2.62)]{Rafaely:2019}.}. This changes the definition of the radial filters $b_n^{-1}(R, \omega)$ because the factor $4\pi\mathrm{i}^{-n}$ now needs to be included into~\eqref{eq:radial_filter_3d} and~\eqref{eq:radial_filter_3d_prime} so that it also gets applied in~\eqref{eq:s_breve_from_s}.\\

\noindent The formulation that has established in the ambisonics community \citep{Daniel:PhD2001,Nachbar:AmbiX2011} employs yet another difference compared to Sec.~\ref{sec:sh_decomposition}. That is, everything in the ambisonics context is formulated in terms of incidence directions rather than propagation directions so that Sec.~\ref{sec:propagation_direction} kicks in. Instead of reformulating the entire Sec.~\ref{sec:sh_decomposition} and~\ref{sec:binaural_rendering}, we remind ourselves of the discussion from Sec.~\ref{sec:definition_of_hrtfs}. We concluded that we can account for a change of the reference direction by replacing the factor $\i^{-n}$, for example in~\eqref{eq:sh_basic_ambi}, with $\i^n$. This changes the radial filters $b_n^{-1}(R, \omega)$ from the inverse of~\eqref{eq:radial_filter_3d} to the inverse of
\begin{equation}\label{eq:radial_filter_3d_2}
    b^{\text{ambi}}_n\!\left( R, \omega\right)  = 4\pi\mathrm{i}^{n}
    \left[ j_n\!\left( \omega \frac{R}{c} \right) - \frac{j_n^{\;\!\prime}\!\left( \omega \frac{R}{c} \right)}{h_n^{\prime\;\!(2)}\!\left( \omega \frac{R}{c} \right)}\,h_n^{(2)}\!\!\left( \omega \frac{R}{c} \right) \right] \ .
\end{equation}
and accordingly for the representation from~\eqref{eq:radial_filter_3d_prime}.
As a consequence, the factors $4\pi\mathrm{i}^{-n}$ or $4\pi\mathrm{i}^{n}$, respectively, need to be removed from the right-hand sides of~\eqref{eq:S_bar_to_breve_S}, \eqref{eq:binaural_rendering_final} and~\eqref{eq:binaural_rendering_final_2}, and the coefficients $\mathring{H}_{n,m}(\omega)$ of the HRTFs need to be computed based on the incidence directions, too. This formulation is compatible with the N3D normalization that is used in the standard ambisonic software tools such as SPARTA\footnote{\url{https://leomccormack.github.io/sparta-site/}} and the IEM Plugin Suite\footnote{\url{https://plugins.iem.at/}}. MATLAB implementations are provided in~\citep{Ahrens:ema_encoding,Ahrens:sma_encoding}.\\

\noindent Refer to Table~\ref{tab:table} for a summary.\\

\noindent As a side note, the coordinate system that is used in ambisonics is a right-hand coordinate system where the $x$-axis points 'forward', the $y$-axis points 'to the left', and the $z$-axis points 'upwards' relative to the reference orientation of a notional listener. 

\section{Conclusions}

We revisited the mathematics of binaural audio rendering in the spherical harmonic domain, and we highlighted the pitfalls, which relate to some of the definitions of the employed quantities. Our findings are summarized in Table~\ref{tab:table} on the subsequent page. We recommend in critical situations, such as when implementing a given method, that the reader uses the present document as reference to verify the consistency of the definitions in the literature that presents the method to be implemented and the definitions used by the implementation. 

We also have to highlight that some of the literature is inconsistent in itself, which includes some of our own publications. 

\setcounter{footnote}{8}

\newgeometry{margin=2cm} 
\begin{landscape}

\begin{table}[ht]
    \centering
    {\footnotesize
    \begin{tabular}{|c|c|c||c|c|}
    \hline
    \vphantom{$\displaystyle{\int^I}$} Definition of & Definition of the & $\,$ Definition of $\,$ & Radial filters  & Binaural rendering\\
    $\breve{S}_{n,m}(\omega)$ & \makecell{temporal Fourier transform\\ --- \\ Reference direction} & $Y_{n,m}(\beta, \alpha)$ & $b_n^{-1}\left( \omega \frac{R}{c}, R \right)$ & \\
    \hline
    \hline
    $\displaystyle S(\vec{x}, \omega) =
    \sum_{n=0}^\infty \sum_{m=-n}^n \breve{S}_{n,m}(\omega)\,\,j_n\!\left( \omega \frac{r}{c} \right)\,Y_{n,m}(\beta, \alpha)$  &  \makecell{Eq.~\eqref{eq:cft} and~\eqref{eq:cift}\\ --- \\ Propagation}  & Eq.~\eqref{eq:sh_definition} & $\left( -
    \frac{\i}{\left(\omega\frac{R}{c}\right)^2} \frac{1}{h_n^{\prime\;\!(2)}\!\left( \omega \frac{R}{c} \right)} \right)^{-1}$  & $\displaystyle B^\text{L,R}(\omega) =\sum_{n=0}^N \sum_{m=-n}^n \!\! \frac{1}{4\pi\i^{-n}} \, \breve{S}_{n,-m}(\omega)\,\,\mathring{H}_{n,m}^\text{L,R}(\omega)$ \\
    \hline
    $\displaystyle S(\vec{x}, \omega) =
    \sum_{n=0}^\infty \sum_{m=-n}^n\!\! 4\pi\i^{-n} \, \breve{S}_{n,m}(\omega)\,\,j_n\!\left( \omega \frac{r}{c} \right)\,Y_{n,m}(\beta, \alpha)$  &  \makecell{Eq.~\eqref{eq:cft} and~\eqref{eq:cift}\\ --- \\ Propagation}   & Eq.~\eqref{eq:sh_definition} & $\left( - 4\pi\i^{-n} \,
    \frac{\i}{\left(\omega\frac{R}{c}\right)^2} \frac{1}{h_n^{\prime\;\!(2)}\!\left( \omega \frac{R}{c} \right)} \right)^{-1}$  & $\displaystyle B^\text{L,R}(\omega) =\sum_{n=0}^N \sum_{m=-n}^n\!\breve{S}_{n,-m}(\omega)\,\,\mathring{H}_{n,m}^\text{L,R}(\omega)$ \\
    \hline
    $\displaystyle S(\vec{x}, \omega) =
    \sum_{n=0}^\infty \sum_{m=-n}^n\!\! 4\pi\i^{-n} \, \breve{S}_{n,m}(\omega)\,\,j_n\!\left( \omega \frac{r}{c} \right)\,Y_{n,m}(\beta, \alpha)$  &  \makecell{Eq.~\eqref{eq:cft} and~\eqref{eq:cift}\\ --- \\ Propagation}   & Eq.~\eqref{eq:sh_definition_2} & $\left( - 4\pi\i^{-n} \,
    \frac{\i}{\left(\omega\frac{R}{c}\right)^2} \frac{1}{h_n^{\prime\;\!(2)}\!\left( \omega \frac{R}{c} \right)} \right)^{-1}$  & $\displaystyle B^\text{L,R}(\omega) =\sum_{n=0}^N \sum_{m=-n}^n\!(-1)^m\,\breve{S}_{n,-m}(\omega)\,\,\mathring{H}_{n,m}^\text{L,R}(\omega)$ \\
    \hline
    $\displaystyle S(\vec{x}, \omega) =
    \sum_{n=0}^\infty \sum_{m=-n}^n\!\! 4\pi\i^{n} \, \breve{S}_{n,m}(\omega)\,\,j_n\!\left( \omega \frac{r}{c} \right)\,Y_{n,m}(\beta, \alpha)$  &  \makecell{Eq.~\eqref{eq:cft} and~\eqref{eq:cift}\\ --- \\ Incidence}    & Eq.~\eqref{eq:sh_definition_real} & $\left( - 4\pi\i^{n} \,
    \frac{\i}{\left(\omega\frac{R}{c}\right)^2} \frac{1}{h_n^{\prime\;\!(2)}\!\left( \omega \frac{R}{c} \right)} \right)^{-1}$  & $\displaystyle B^\text{L,R}(\omega) =\sum_{n=0}^N \sum_{m=-n}^n\!\breve{S}_{n,m}(\omega)\,\,\mathring{H}_{n,m}^\text{L,R}(\omega)$ \\
    \hline
    $\displaystyle S(\vec{x}, \omega) =
    \sum_{n=0}^\infty \sum_{m=-n}^n\!\! 4\pi\i^{-n} \, \breve{S}_{n,m}(\omega)\,\,j_n\!\left( \omega \frac{r}{c} \right)\,Y_{n,m}(\beta, \alpha)$  & \makecell{Eq.~\eqref{eq:cft} and~\eqref{eq:cift} but with the sign \\ of the exponent swapped\\ --- \\ Incidence}  & Eq.~\eqref{eq:sh_definition_real} & $\left( 4\pi\i^{-n} \,
    \frac{\i}{\left(\omega\frac{R}{c}\right)^2} \frac{1}{h_n^{\prime\;\!(1)}\!\left( \omega \frac{R}{c} \right)} \right)^{-1}$  & $\displaystyle B^\text{L,R}(\omega) =\sum_{n=0}^N \sum_{m=-n}^n\!\breve{S}_{n,m}(\omega)\,\,\mathring{H}_{n,m}^\text{L,R}(\omega)$ \\
    \hline
    \end{tabular}
    \caption{Five typical combinations of definitions. The right two columns are a consequence of the definitions in the left three columns. The fourth and fifth rows both yield ambisonic signals $\breve{S}_{n,m}(\omega)$ that correspond to the N3D ambisonic standard\protect\footnotemark.}
    \label{tab:table}
    }
\end{table}

\footnotetext{See \url{https://github.com/AppliedAcousticsChalmers/ambisonic-encoding}}

\end{landscape}
\restoregeometry

\end{document}